\documentstyle[12pt,worldsci]{article}
\input epsf
\begin{document}
\vbox{\hfill\hbox{JLAB-THY 99--19}}

\title{QCD Constraints on Form Factor Shapes\footnote{Presented at
``Exclusive \& Semi-exclusive Processes at High Momentum Transfer,''
Jefferson Lab, Newport News, VA, May 20--22, 1999; to appear in
Proceedings.}\footnote{Work supported by the Department of Energy
under contract No.\ DE-AC05-84ER40150.}}

\author{Richard F. Lebed\\
{\em Jefferson Lab, Newport News, Virginia 23606, USA}}
\maketitle
\setlength{\baselineskip}{2.6ex}

\vspace{0.7cm}
\begin{abstract}

	This talk presents an introduction to the use of dispersion
relations to constrain the shapes of hadronic form factors consistent
with QCD.  The applications described include methods for studying
$|V_{cb}|$ and $|V_{ub}|$, the strange quark mass, and the pion charge
radius.

\end{abstract}
\vspace{0.7cm}

\section{Introduction and History}

	Between the mid 1950s and late 1960s, a great deal of
theoretical activity focused on attempting to solve (or at least
severely constrain) problems of strong interaction physics using
dispersion theory.  An extensive and elegant body of work was
developed to study the analyticity properties of form factors and
scattering amplitudes.  Eventually, however, when theorists believed
they had reached the the limits of what could be gleaned from
dispersive techniques, their attentions were drawn elsewhere: to the
quark-parton model, to current algebra, and eventually to gauge
theories, especially QCD.

	The appeal of dispersion theory lies in its ability to
incorporate in a completely rigorous and model-independent fashion
those features shared by all well-defined field theories, namely,
causality, unitarity, and crossing symmetry.  Moreover, it works
equally well in perturbative and nonperturbative regimes of the
underlying dynamical theory.  However, no specific Lagrangian is
demanded by this scheme, and this lack of specificity acts as a
double-edged sword: Without dynamical input, one can only deduce those
consequences common to all possible dynamics.  On the other hand, we
now possess QCD, which is the fundamental, albeit unsolved, theory of
strong interactions.  A combination of the two, in which QCD inputs
are inserted directly into dispersion relations, should yield a rich
harvest of rigorous and model-independent bounds on hadronic
quantities.  In this talk we explore the implementation of this idea
to the specific cases of weak and electromagnetic hadronic form
factors.

	Dispersion theory has been with us in particle physics for
quite some time.  The origin of the name traces directly back to the
famous Kramers-Kronig relation\cite{KK} in electromagnetism, used to
describe the dispersion of light in an arbitrary medium.  The standard
formula reads
\begin{equation}
{\rm Re} \, f(\omega) = {\rm Re} \, f(0) + \frac{\omega^2}{2\pi^2} P
\int_0^\infty d\omega^\prime \frac{\sigma_{\rm tot}
(\omega^\prime)}{\omega^{\prime 2} - \omega^2} ,
\end{equation}
where $f$, $\omega$, $\sigma_{\rm tot}$, and $P$ denote, respectively,
the forward scattering amplitude, frequency, total cross section, and
principal value prescription to remove the denominator singularity.
It follows that the dispersive (Re$\, f$) and absorptive (Im$\, f \sim
\sigma_{\rm tot}$) amplitudes are intimately connected.  This relation
is derived by using causality and unitarity, which lead to
restrictions on the analyticity properties of $f$ in the space of
complex $\omega$.  From there, the elegant theorems of complex
analysis, especially Cauchy's theorem, provide the identities known as
dispersion relations.

	Since quantum mechanics and quantum field theory are expressed
over the field of complex numbers, it is natural to expect that some
variant of the dispersive approach should also exist in particle
theory.  Indeed, as early as 1951, Gell-Mann, Goldberger, and
Thirring\cite{GGT} described how causality and unitarity lead to a
dispersion relation for the vacuum polarization two-point function of
QED.  A flurry of other dispersion relations followed in the
literature, each presented with more or less rigor, depending upon
assumptions about the analytic structure of the quantity under
scrutiny; however, the particular dispersion relation used below is
nothing more than the QCD version of the one first studied in 1951.

\vspace{0.1cm}

\section{Formalism}

	We begin by defining the vacuum polarization tensor as the
two-point current correlator in momentum space:
\begin{equation}
\Pi^{\mu\nu} (q) \equiv i \int d^4 x \, e^{iqx} \langle 0 | {\rm T} J^\mu
(x) J^{\dagger\nu} (0) | 0 \rangle ,
\end{equation}
Here $J$ is some chosen current; since we are working with QCD, we
choose it to be a quark bilinear.  Moreover, we choose $J$ to be a
weak or electromagnetic, rather than gluonic, current, so that the
individual (perturbative) current insertions are easier to identify.
Suppressing for now the Lorentz indices $\mu \nu$, we would like to
use Cauchy's theorem to write an expression
\begin{equation}
\Pi (q^2) = \frac{1}{2\pi i} \int_C dt \, \frac{\Pi (t)}{(t-q^2)} ,
\end{equation}
which relates $\Pi$ at two different momentum arguments, $q^2$ and
$t$.  However, in order to do this, we must identify a closed contour
$C$, inside of which $\Pi$ is analytic in $t$.  In the present case,
causality implies that $\Pi(t)$ is analytic in $t$ except on parts of
the positive $t$ axis, where $J^\dagger$ can create on-shell hadrons,
which generates a discontinuity only in the imaginary (absorptive)
part of $\Pi$.  We choose $C$ to consist of the lower and upper sides
of this branch cut, together with the circle with $|t| \to \infty$;
the latter contribution vanishes as long as $\Pi \to 0$ for large
$|t|$.  Then
\begin{equation} \label{cont}
\Pi (q^2) = \frac{1}{2\pi} \int_0^\infty dt \, \frac{{\rm Im} \, \Pi
(t+i\epsilon)} {(t-q^2)} + \frac{1}{2\pi} \int_\infty^0 dt \,
\frac{{\rm Im} \, \Pi (t-i\epsilon)} {(t-q^2)} .
\end{equation}
Using the Schwarz reflection principle ($\Pi(z^*) = \Pi^* (z)$ if
$\Pi$ is real on some segment of the real axis, which is true for
$t<0$ since there are no on-shell thresholds and hence is no imaginary
part there), the two terms in (\ref{cont}) are equal:
\begin{equation}
\Pi(q^2) = \frac{1}{\pi} \int_0^\infty dt \, \frac{{\rm Im} \, \Pi
(t+i\epsilon)} {(t-q^2)} .
\end{equation}
If $\Pi(q^2)$ diverges, or the contribution from the circle $|t| \to
\infty$ does not vanish, such terms may be removed through the process
called ``subtraction'': Since the offending terms appear as
coefficients of a polynomial in $q^2$, taking a sufficient number $n$
of $q^2$ derivatives yields a finite result,
\begin{equation}
\frac{\partial^n \Pi (q^2)} {\left( \partial q^2 \right)^n} =
\frac{\partial^n \Pi_{\rm finite} (q^2)} {\left( \partial q^2
\right)^n} .
\end{equation}
Then the expression for the dispersion relation reads
\begin{equation} \label{d1}
\Pi^{(n)} (q^2) \equiv \frac{1}{n!} \frac{\partial^n \Pi (q^2)}
{\left( \partial q^2 \right)^n} = \frac{1}{\pi} \int_0^\infty dt \,
\frac{{\rm Im} \, \Pi (t+i\epsilon)} {(t-q^2)^{n+1}} .
\end{equation}
Restoring the Lorentz indices and inserting a complete set of states
between $J$ and $J^\dagger$ (unitarity) yields
\begin{equation} \label{unit}
{\rm Im} \, \Pi^{\mu\nu} (t+i\epsilon) = \frac 1 2 \sum_{\Gamma} \int
d\Phi(\Gamma) (2\pi)^4 \delta^4 \left( t-\sum_\Gamma p \right) \langle
0 | J^\mu | \Gamma \rangle \langle \Gamma | J^{\dagger\nu} | 0 \rangle
,
\end{equation}
where only on-shell states $\Gamma$ with phase space $\Phi$ are
included in the sum (a consequence of reducing the step functions in
the time ordering).  The matrix elements $\langle 0|J^\mu| \Gamma
\rangle$ are nothing more than decay constants and form factors---pure
hadronic quantities---while $q^2$ can be chosen so that $\Pi(q^2)$ can
be evaluated directly in the fundamental theory of QCD.  In
particular, one chooses $q^2$ to be far from the hadronic (strong
coupling) region, and then $\Pi(q^2)$ may be computed using an
operator product expansion.  A very useful observation due to
Meiman\cite{Meim} in 1963 is that the $\mu=\nu$ components of
Eq.~(\ref{unit}) are positive definite, meaning that each hadronic
contribution serves only to saturate further the partonic
(perturbative QCD) side of the dispersion relation.  In this way one
obtains a rigorous inequality between partonic and hadronic physics.

	One path from Eqs.~(\ref{d1})--(\ref{unit}) leads to the
famous QCD sum rules\cite{sumrule}, which study the saturation of the
equality between the partonic and hadronic sides.  We focus also on
what this equality tells us about the behavior of matrix elements
$\langle 0|J^\mu| \Gamma \rangle$.  The first work to use the Meiman
inequality with QCD inputs was by Bourrely, Machet, and de
Rafael\cite{BMR} in 1981.

	As an explicit example, consider the pion electromagnetic form
factor:
\begin{equation}
\langle \pi^+ (p^\prime) | J^\mu_{\rm EM} | \pi^+ (p) \rangle = f(q^2)
(p+p^\prime)^\mu ,
\end{equation}
where $q = p - p^\prime$.  Then Eq.~(\ref{unit}) becomes
\begin{equation}
{\rm Im} \, \Pi^{ii} (t) \ge \frac{1}{48\pi} \left( t-4m_\pi^2
\right)^{3/2} t^{-1/2} |f(t)|^2 \theta ( t-4m_\pi^2) ,
\end{equation}
while the partonic side, finite after two subtractions, is computed to
be
\begin{equation}
\Pi^{ii \, (2)} (q^2) = \frac{1}{8\pi(-q^2)} \left\{ 1 +
\frac{\alpha_s (q^2)}{\pi} + O \left[ \left( \frac{\alpha_s^2
(q^2)}{\pi} \right)^2 \right] + n.p. \right\} ,
\end{equation}
where $n.p.$ stands for nonperturbative corrections such as vacuum
condensates.  The combined inequality reads
\begin{equation}
\frac{1}{8\pi(-q^2)} \left\{ 1 + \frac{\alpha_s (q^2)}{\pi} + O \left[
\left( \frac{\alpha_s^2 (q^2)}{\pi} \right)^2 \right] + n.p. \right\}
\ge \frac{1}{48\pi^2} \int_{4m_\pi^2}^\infty dt \,
\frac{(t-4m_\pi^2)^{3/2}}{t^{1/2} (t-q^2)^3} |f(t)|^2 .
\end{equation}

	In general, one obtains an inequality of the form
\begin{equation}
\frac{1}{\pi} \int_{t_+}^\infty dt \, \frac{W_F (t)
|F(t)|^2}{(t-q^2)^{n+1}} \le \Pi^{(n)} (q^2) ,
\end{equation}
where $t_+$ is the lowest threshold and $W_F$ is a positive weighting
factor arising from phase space and the quantum numbers of the form
factor $F(t)$.  As discussed by Okubo and Fushih\cite{Okubo} in 1971,
it is very convenient to map the complex $t$ plane with a cut for $t_+
\le t < +\infty$ to the unit disc using a complex kinematic variable
$z$:
\begin{equation}
z(t;t_s) \equiv \frac{\sqrt{t_+ - t} - \sqrt{t_+ - t_s}} {\sqrt{t_+ -
t} + \sqrt{t_+ - t_s}} ,
\end{equation}
which maps the upper (lower) side of the cut to the lower (upper) half
of the unit circle.  The parameter\cite{BL} $t_s < t_+$ is chosen
later for convenience.  One then defines a weighting function
\begin{equation}
\phi_F (t;t_s) = \sqrt{\frac{W_F(t)}{\Pi^{(n)} (q^2) (t-q^2)^{n+1}
|dz(t;t_s)/dt|}} ,
\end{equation}
which is analytic inside the unit circle, and in terms of which the
dispersive bound reads
\begin{equation}
\frac{1}{2\pi i} \int \hspace{-1.1em} \bigcirc \frac{dz}{z} \left|
\phi_F (z) F(z) \right|^2 \le 1 .
\end{equation}
If any subthreshold poles remain inside the unit circle at points
$z=z_p$, they may be removed by means of so-called Blaschke factors,
\begin{equation}
P_F (z) = \prod_p \frac{z-z_p}{1- z_p^* z} .
\end{equation}
$P_F (z)$ has the feature that for $|z|=1$, $|P_F(z)| = 1$, so that
the dispersive bound is unchanged,
\begin{equation} \label{bound}
\frac{1}{2\pi i} \int \hspace{-1.1em} \bigcirc \frac{dz}{z} \left|
\phi_F (z) P_F (z) F(z) \right|^2 \le 1 ,
\end{equation}
and $\phi_F (z) P_F (z) F(z)$ is analytic on the whole unit disc.
Crossing symmetry relates the form factor in all kinematic regimes by
analytic continuation.  We have thus isolated the analytic structure
of the form factor,
\begin{equation} \label{analyt}
F(t) = \frac{1}{P_F(t) \phi_F (t;t_s)} \sum_{n=0}^\infty a_n z^n
(t;t_s) ,
\end{equation}
where the coefficients $a_n$ are unknown; however, inserting
Eq.~(\ref{analyt}) back into (\ref{bound}) gives
\begin{equation} \label{an}
\sum_{n=0}^\infty | a_n |^2 \le 1 .
\end{equation}
Equations (\ref{analyt}) and (\ref{an}) first appeared\cite{BGL1} in
1995, and re-express in a very compact and explicit notation all of
the analyticity, unitarity, and explicit QCD information implicit in
Eqs.~(\ref{d1}) and (\ref{unit}).  Since $\phi_F$ and $P_F$ are known
functions, the form factor is known except for a set of parameters
$a_n$, each of which must be less than unity in magnitude.  A randomly
chosen shape for a form factor would almost inevitably have some
$|a_n| > 1$, and thus would be disallowed by the dispersive bound
(\ref{an}).

	One more point that makes the model-independent
parameterization (\ref{analyt}) useful is that for spacelike and
semileptonic processes, the allowed kinematic range for $z$ tends to
have $|z| \ll 1$.  Indeed, the parameter $t_s$ is chosen to enhance
this effect.  For example, for $\bar B \to D \ell \bar \nu$, $|z| <
0.03$.  This means that the convergence of Eq.~(\ref{analyt}) is
geometrically fast, and only the first few $a_n$'s are relevant to the
shape of the form factor.  The theoretical uncertainty incurred by
ignoring the other, infinite set of $a_n$'s is called ``truncation
error,''\cite{BGL1} and falls off geometrically fast with the number
of $a_n$'s used to parameterize the form factor.

\vspace{0.1cm}

\section{A Gallery of Results}

\begin{figure} \label{fig1}
  \begin{centering}
        \def\epsfsize#1#2{1.10#2}
        \vbox{\hfil\epsfbox{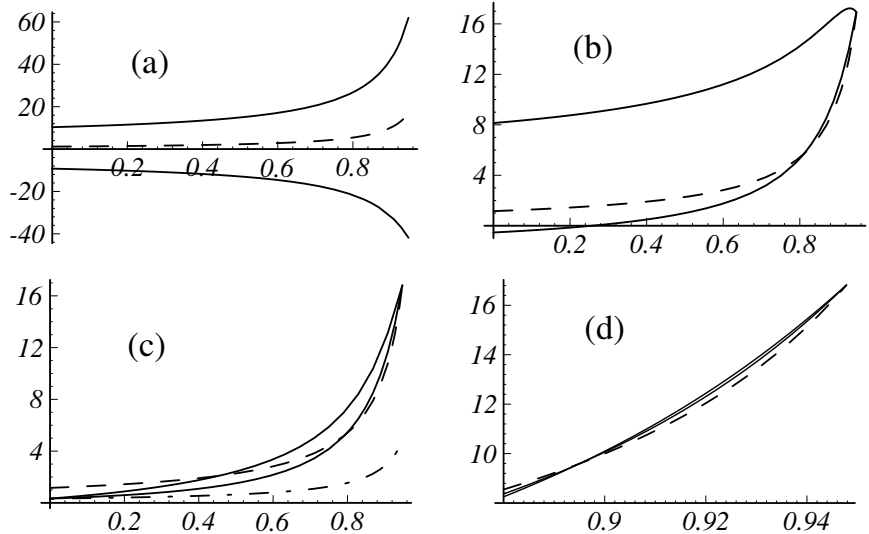}\hfill}
%        \epsfbox{first1.eps}\hfil}

\caption{Bounds on the $\bar B \to \pi \ell \bar \nu$ form factor $f$
using the dispersive bound and fixing zero, one, and two points in
$(a)$, $(b)$, and $(c)$, respectively. Dashed lines indicate pole
dominance models.  $(d)$ shows how certain choices of $B^*$ pole
parameters can violate the dispersive bounds.}

  \end{centering}
\end{figure}

	1) {\it $|V_{ub}|$ and $|V_{cb}|$.}  The need for a
parameterization describing all solutions of
Eqs.~(\ref{d1})--(\ref{unit}) was recognized\cite{BGLprl} in studies
of the $\bar B \to \pi \ell \bar \nu$ form factor, useful for the
extraction of $|V_{ub}|$.  There it was seen that the inclusion of
each (at that time, hypothetical) form factor data point served to
decrease the region allowed by the dispersion relation geometrically
fast (Fig.~\ref{fig1}).  Similar comments apply to using points from a
lattice simulation.\cite{Lel}

	The model-independent form factor parameterization
Eq.~(\ref{analyt}) was first used\cite{BGL1} to extrapolate measured
$\bar B \to D^{(*)} \ell \bar \nu$ form factor data to a point where
phase space vanishes.  In order to extract $|V_{cb}|$ from the form
factor\footnote{Strictly speaking, 4 form factors contribute; however,
in the limit of heavy quarks, each one either vanishes or is
proportional to a single ``Isgur-Wise'' form factor\cite{IW}.} in the
differential width
\begin{equation} \label{par}
\frac{d\Gamma}{dq^2} ( \bar B \to D^* \ell \bar \nu ) \propto \left|
V_{cb} \right|^2 \left| F(q^2) \right|^2 \sqrt{(M_B + M_{D^*})^2 -
q^2} ,
\end{equation}
one must separate $|V_{cb}|$ from $|F(q^2)|$.  The normalization of
$F$, namely, $F(q^2=(M_B-M_{D^*})^2) = 1$ (up to small corrections),
is determined by the heavy quark limit\cite{IW}.  However,
Eq.~(\ref{par}) shows that phase space vanishes at exactly this $q^2$;
therefore, an extrapolation is needed.  Previously, experimental
measurements of the form factor used an {\it ad hoc\/} linear or
quadratic extrapolation, which implies a theoretical uncertainty of
unknown size (see Fig.~2$a$).  Using (\ref{analyt}) and (\ref{an})
removes this uncertainty, and subsequent work\cite{BDlater,BDfits}
refined the analysis to the point that it is used by both theorists
and the experimental groups themselves (Table~\ref{vcb}).
\begin{figure} \label{fig2and3}
  \begin{centering}
        \def\epsfsize#1#2{0.38#2}
        \vbox{\hfil\epsfbox{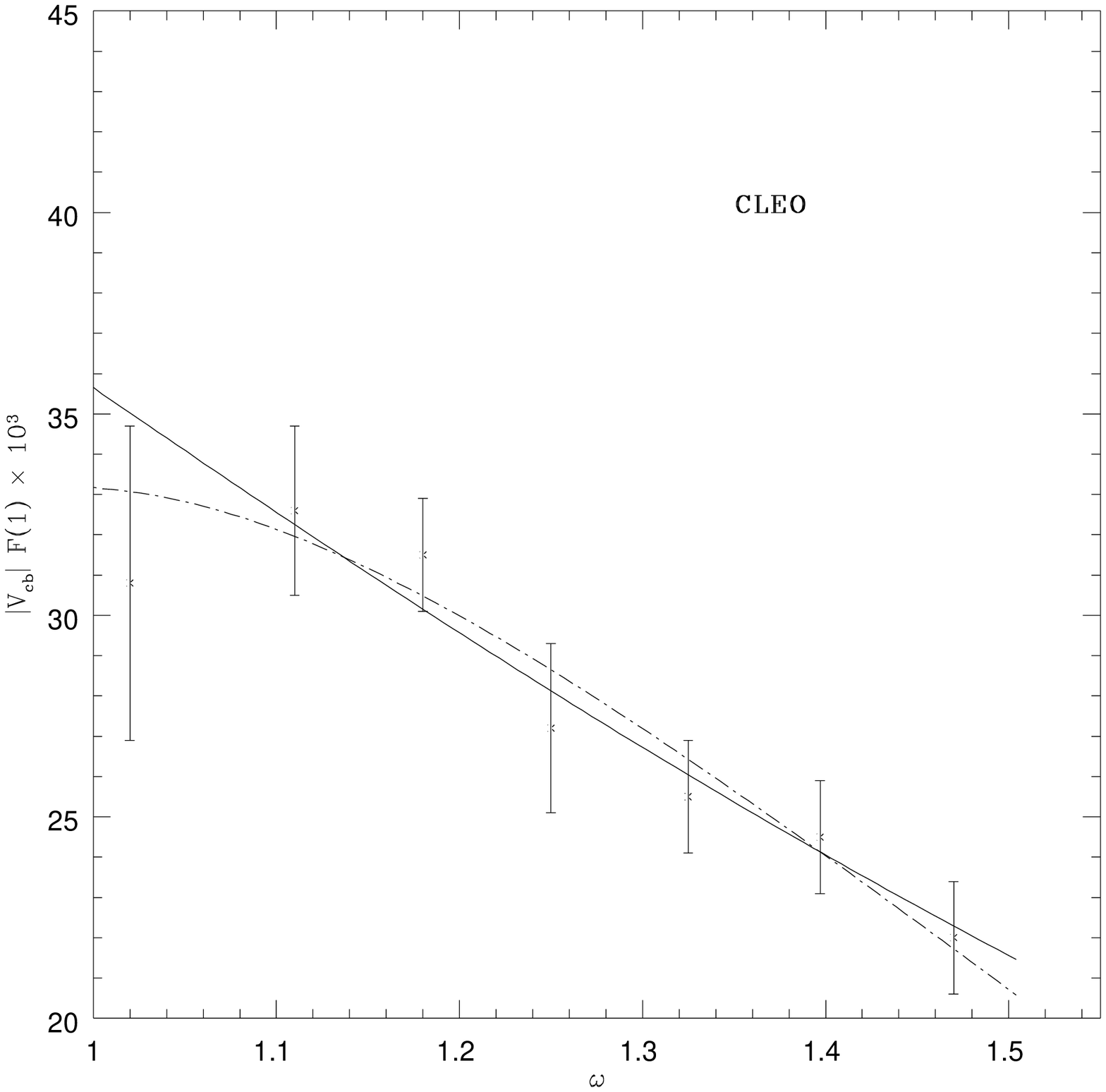}\hfill
        \epsfbox{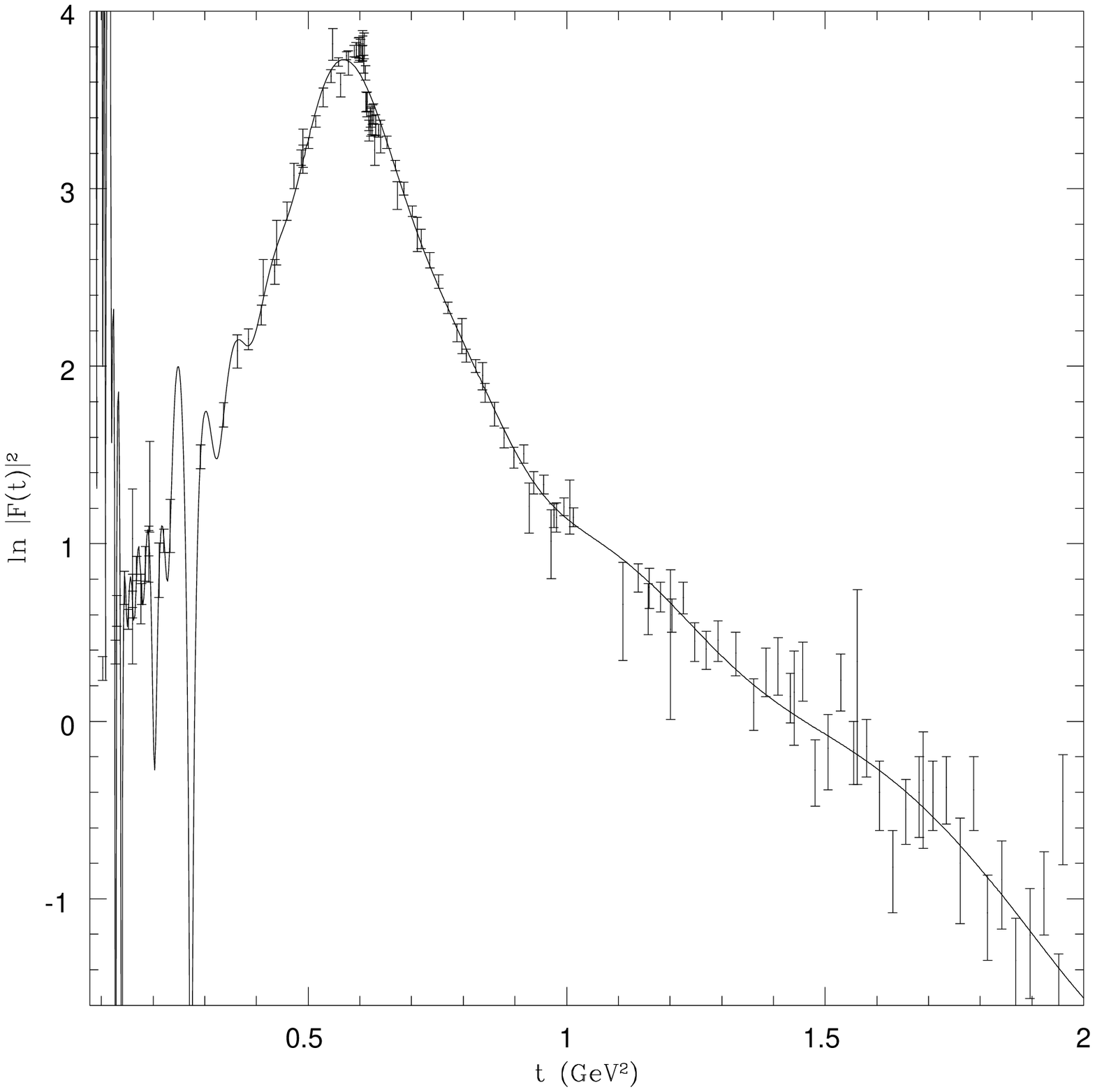}\hfil}

\caption{$(a)$.  Fit to CLEO data for the $\bar B \to D^* \ell \bar
\nu$ form factor using the model-independent parameterization
(\ref{analyt}) with (solid) and without (dashed) the dispersive
constraint Eq.~(\ref{an}).  The eye prefers the dashed curve, but it
is forbidden by QCD.  $(b)$.  Fit to timelike pion electromagnetic
form factor data.  Contrast the smoothness of the fit above $t=0.4$
GeV$^2$ to the oscillations below.}

  \end{centering}
\end{figure}
\begin{table}
\hfil\begin{tabular}{c|c|c}
Experiment & Process & $|V_{cb}| \cdot 10^3$ \\ \hline\hline

CLEO\cite{BDfits} & $\bar B \to D^* \ell \bar \nu$ &
$36.9^{+2.0}_{-2.1}$ \\

CLEO\cite{CLEOfit} & $\bar B \to D \ell \bar \nu$ & $44.8 \pm 6.1$ \\

ALEPH\cite{BDfits} & $\bar B \to D^* \ell \bar \nu$ & $31.9 \pm 2.4$
\\

ALEPH\cite{BDfits} & $\bar B \to D \ell \bar \nu$ & $29.2 \pm 7.3$ \\

DELPHI\cite{DELPHI} & $\bar B \to D^* \ell \bar \nu$ & $38.0 \pm 1.3$
\\ \hline

\end{tabular} \hfil
\caption{Determinations of $|V_{cb}|$ using Eqs.~(\ref{analyt}) and
(\ref{an}).  Footnotes reference the source of the fit.  Uncertainties
are statistical plus theoretical.\label{vcb}}
\end{table}

	2) {\it Strange quark mass.}  $K_{\ell 3}$ decays possess two
form factors, one of which appears with the coefficient $m_{\ell}^2$
in the rate and is called the scalar form factor.  The corresponding
$\Pi^{(n)} (q^2)$, evaluated deep in the Euclidean region, is
proportional to $(m_s-m_u)^2$, {\it i.e.}, is sensitive to $m_s$.  One
can invert the program of 1), so that a large amount of form factor
data, thus delineating its shape, is used\cite{LS} to constrain the
function $\Pi^{(n)} (q^2)$ and hence $m_s$.  Indeed, one finds that
$F(t) \propto m_s a_n$, so that (\ref{an}) implies a rigorous lower
bound on $m_s$.

	Currently, not enough data exists in the world sample for such
a determination, although DA$\Phi$NE expects to increase the available
pool many times over.  Until such data exists, one may apply the
results of a model, or better, a chiral perturbation theory ($\chi$PT)
calculation to compute the required shape parameters $a_n$.  Believing
$\chi$PT to a level of 5\%, 1\%, or 1/20\% leads to lower bounds on
$m_S^{\overline{\rm MS}}$(1 GeV) of 40, 90, and 140 MeV, respectively.

	3) {\it Pion form factor.} The parameterization (\ref{analyt})
exhibits geometric convergence for $t<t_+$ ($|z|<1$).  On the other
hand, one often possesses data directly on the cut $t>t_+$ ($|z|=1$),
the timelike region.  Does (\ref{analyt}) have anything to say about
this region?  Although one must be much more careful about
convergence, the answer\cite{BuL} appears to be yes.  Theorems of
complex convergence plus knowledge of asymptotic $(t \to \infty)$
properties of form factors allow one to use (\ref{analyt}) even in the
timelike region.

	For example, for the pion electromagnetic form factor, one
obtains the fit of Fig.~2$b$.  The presence of the $\rho$ peak is not
put in by hand, but simply emerges from fitting to (\ref{analyt}).
Note, however, the wild oscillations for $t < 0.4$ GeV$^2$; one can
show that these occur due to large gaps in the data for $\theta >
\pi/2$ on the unit circle $|z|=1$.  These large oscillations persist
when one analytically continues into the spacelike region, where one
obtains
\begin{equation}
|F (t=0)| = 2.56 \pm 2.00, \ \ \langle r^2 \rangle = 2.66 \pm 3.44 \,
 {\rm fm}^2 ,
\end{equation}
which are rather loose bounds, considering that, {\it e.g.},
$|F(t=0)|=1$ by charge conservation.  This points to the well-known
problem of the instability of analytic continuation of discrete
timelike data to the spacelike region; now, however one can quantify
exactly how unstable this continuation is.

	One can also proceed directly with spacelike pion form factor
data alone, where the geometric convergence of (\ref{analyt}) is
restored.  Then one finds,\cite{BuL} using this
model-in\-de\-pen\-dent parameterization, $\langle r^2 \rangle = 0.480
\pm 0.020$ fm$^2$, a few $\sigma$ larger than the usual numbers quoted
in the literature ($\simeq 0.42$ fm$^2$), which use {\it ad hoc}
parameterizations.

\vspace{0.1cm}

\section{Conclusions}

	Dispersive techniques provide an elegant, rigorous bridge
between hadronic and elementary quantities.  For semileptonic or
electromagnetic decays, they provide a model-independent, rapidly
convergent parameterization of form factors.  We have seen that a
number of different problems have already been studied using this
method.  Nucleon form factors, the $K$ charge radius, and improvement
of timelike form factors are obvious future directions.  The reader
can doubtless imagine many others.

\end{document}